\documentclass[pra,twocolumn,tightenlines,showpacs,floatfix]{revtex4}

\usepackage{amsmath}
\usepackage{graphicx}
\usepackage{amssymb}
\usepackage{color}

\renewcommand{\vec}[1]{\boldsymbol #1}

\DeclareMathOperator{\Trace}{Tr}
\DeclareMathOperator{\STr}{STr}

\begin{document}

\title{Renormalization group, dimer-dimer scattering, and three-body forces}

\author{Boris Krippa$^{1,2}$, Niels R. Walet$^{2}$, Michael C. Birse$^{2}$}

\affiliation{$^{1}$Institute for Theoretical and Experimental Physics, Moscow,
117259, Russia\\
$^{2}$School of Physics and Astronomy, The University of Manchester,
Manchester, M13 9PL, UK}

\date{\today}
\begin{abstract}
We study the ratio between the fermion-fermion scattering length and the
dimer-dimer scattering length for systems of nonrelativistic fermions, 
using the same functional renormalisation technique as previously applied 
to fermionic matter. We find a strong dependence on the cutoff function
used in the renormalisation flow for a two-body truncation of the action. 
Adding a simple three-body term substantially reduces this dependence. 
\end{abstract}


\maketitle

Ultra-cold Fermi gases provide a fertile ground for research in
atomic physics. An important feature of such systems is superfluidity,
which is the result of attractive fermion-fermion interactions leading
to pairing. Recent advances using Feshbach resonances allow a tuning
of the fermion-fermion S-wave scattering length $a_{F}$. For negative 
scattering length we get the weak-coupling BCS state. For positive values 
of $a_{F}$ bound states of two fermions--{}``dimers''--form and these
can lead to a Bose-Einstein condensate (BEC) \cite{Zwe}. The
size of dimers is determined by the fermion-fermion scattering length
and their binding energy is of order $1/a_{F}^{2}$.

For a sufficiently dilute and cold gas of dimers the main dynamical
quantity characterising their interaction is the dimer-dimer scattering
length $a_{B}$. The exact relation between dimer-dimer and fermion-fermion
scattering lengths $a_{B}=0.6a_{F}$ was established in Ref.~\cite{Shl}
by solving the Schr\"odinger equation for two composite bosons interacting
with an attractive zero-range potential. This method is difficult
to extend to the many-body case. Therefore, it is useful to study
the ratio $a_{B}/a_{F}$ in an approach which can be used both for
few and many-body problems.

In this paper we calculate $a_{B}$ in the framework of the Exact
Renormalisation Group (ERG) approach. (For reviews, see 
Refs.~\cite{BTW,DMT}). This technique has been previously used to 
study a variety of physical systems, from systems of nonrelativistic 
fermions \cite{Kri,Diehl,Bir,DKS,Flo,Flo2} to quark models \cite{JW} and 
gauge theories \cite{LPG}. It is based on the scale-dependent average 
effective action $\Gamma_{k}$, where $k$ is an auxiliary running scale. 
The action at scale $k$ contains the effects of field fluctuations 
with momenta $q$ larger than $k$ only. In the limit $k\rightarrow0$ 
all fluctuations are included and the full effective action is recovered. 
In practice one introduces a set of $k$-dependent cutoff functions $R(q)$, 
which suppress the effect of modes with $q<k$ in the path integral for 
the action by giving them a large $k$ dependent mass. The functions
$R(q)$ should vanish in the limit $k\rightarrow0$ and behave like 
$k^{\alpha}$ with $\alpha>0$ when $q\rightarrow0$. 

With this prescription the average effective action at large $k$
is just the classical action of the theory---in our case  
nonrelativistic fermions with a local interaction. An exact 
solution of the functional RG equation should be independent of
the choice of cutoff for $k\rightarrow 0$. However, in practice, 
truncations of the action inevitably lead to some cutoff dependence
of the results. We can use this dependence as a measure of the quality 
of the truncation. With this tool, we shall see that the standard 
parametrisation of the effective action containing only two-body 
terms is insufficient, and that we get better results by including 
the simplest three-body term. 

The flow of the effective action satisfies \cite{BTW}
\begin{equation}
\partial_{k}\Gamma=-\frac{i}{2}\,\STr\left[(\partial_{k}R)\,
(\Gamma^{(2)}-R)^{-1}\right].\label{eq:ERG1}\end{equation}
where $\Gamma^{(2)}$ is the second functional derivative with
respect to the fields, and the cutoff functions in the mass-like term 
$R(k)$ drive the RG evolution.
The operation $\STr$ denotes the supertrace \cite{Ma} taken over
energy-momentum variables and internal indices and is defined by 
\begin{equation}
\STr\left(\begin{array}{cc}
A_{BB} & A_{BF}\\
A_{FB} & A_{FF}\end{array}\right)=\Trace(A_{BB})-\Trace(A_{FF}).\end{equation}

The evolution equation for the average effective action has a one-loop structure, 
but contains a fully dressed, scale-dependent propagator 
$(\Gamma^{(2)}-R)^{-1}$. Thus, despite its apparently simple form,  
Eq.~(\ref{eq:ERG1}) is actually a functional differential equation. 
In the absence of general methods to solve such equations numerically we must 
resort to approximations. One common approach is to parametrise the effective 
action with a finite set of terms, turning the evolution into a system of 
coupled ordinary differential equations for their coefficients. These 
equations can then be solved numerically. Here we study possible truncations for 
fermionic few-body systems, and our choice of ansatz for the action is motivated 
by both ERG studies of many-body systems \cite{Kri,Diehl} and effective field 
theories (EFTs) for few-fermion systems \cite{EFT}. 
The technique we use is similar to the one used in Ref.~\cite{Bir}   
to analyse the scattering in the system of two nonrelativistic fermions.

A rather different approach to the low-energy fermion-dimer scattering
was considered in Ref.~\cite{DKS}. There the Skorniakov-Ter-Matirosian
equation \cite{STM,EFT} was derived from the RG flow with energy- and
momentum-dependent three-body couplings. Whilst the results obtained
in that work demonstrate the formal equivalence between an RG approach
and three-body quantum mechanics, it seems to be very difficult to
extend the treatment to the more complicated cases of four-body or
many-body systems. In the present paper we focus on the effect of the
three-body forces in the system of two dimers, keeping in mind
possible extensions of the formalism to many-body systems, such as
that attempted in Ref.~\cite{Flo2}. Therefore we only include the
simplest three-body term.

The formation of the trion -- the correlated state of three fermions --
was also studied in Refs.~\cite{Flo,Mor}, but the effect of the three-body 
interactions in the context of the four-body systems was not considered.

We first examine the case when fermions interact only pairwise.
This has previously been considered in Refs.~\cite{Diehl,Bir}, where many 
the technical details can be found, and so we give only a brief account 
of the formalism, concentrating on the dimer-dimer scattering length $a_B$. 
We use this to extend the results of Ref.~\cite{Diehl} and examine the 
cutoff dependence of $a_B$. We then turn to our main task, the inclusion
of three-body terms in $\Gamma$.

As the cutoff scale tends to infinity, we demand the action to be a purely 
fermionic one containing a contact two-body interaction without derivatives.
This kind of interaction has been extensively used in the EFT-based
studies of nuclear forces \cite{EFT}. It is convenient to re-express
the theory in terms of an auxiliary composite boson field by making a
Hubbard-Stratonovich transformation. This replaces the two-body interaction
by a Yukawa-type coupling between the fermions and the auxiliary boson. 
A kinetic term for the boson is then generated by the RG evolution. 
The minimal effective action used in previous work is \cite{Kri,Diehl,Bir}
\begin{align}
&\Gamma_{\mathrm{min}}[\psi,\psi^{\dagger},\phi,\phi^{\dagger},k]\nonumber\\
 & =\int d^{4}x\,\Biggl[\int d^{4}x'\,\phi^{\dagger}(x)\Pi(x,x';k)\phi(x')\nonumber \\
 & \quad+\psi^{\dagger}(x)\left(i\partial_{t}+\frac{1}{2M}\,\nabla^{2}\right)\psi(x)\nonumber \\
 & \quad-\frac{i}{2}\,g\biggl(\psi^{{\rm T}}(x)\sigma_{2}\psi(x)\phi^{\dagger}(x)
-\psi^{\dagger}(x)\sigma_{2}\psi^{\dagger{\rm T}}(x)\phi(x)\biggr)\Biggr]\nonumber\\
&\quad-\frac{1}{2}\,u_2\,\Bigl(\phi^{\dagger}(x)\phi(x)\Bigr)^2.\label{eq:ansatz-1}
\end{align}
Here $\Pi(x,x',k)$ is the scale-dependent boson self-energy and
$u_2$ parametrises the boson-boson interaction which can be generated by
the evolution. The latter is equivalent to a four-body interaction in
terms of the underlying fermions. To this action we add a local three-body interaction, similar to that used in another context in Ref.~\cite{Flo}
Expressed in terms of the boson field this has the form
\begin{eqnarray}
\Gamma[\psi,\psi^{\dagger},\phi,\phi^{\dagger},k]&=&
\Gamma_{\mathrm{min}}[\psi,\psi^{\dagger},\phi,\phi^{\dagger},k]\cr
&&-\lambda\int d^{4}x\,\psi^{\dagger}(x)\phi^{\dagger}(x)\phi(x)\psi(x).\cr
&& \label{eq:ansatz3}\end{eqnarray}

We concentrate first on the two-body part of Eq.~(\ref{eq:ansatz-1}).
The evolution of the boson self-energy is given by 
\begin{equation}
\partial_{k}\Pi(x,x',k)=\frac{\delta^{2}}{\delta\phi(x')\delta\phi^{\dagger}(x)}\,
\partial_{k}\Gamma|_{\phi=0},\end{equation}
although from now on we shall express all evolution in momentum space.
Note that only fermion loops contribute to the evolution of the boson
self-energy in vacuum. These depend on the fermionic cutoff function
$R_{F}$, for which we take the form \cite{Litim} 
\begin{equation}
R_{F}({\vec{q}},k)=\frac{k^{2}-q^{2}}{2M}\,\theta(k-q).\label{eq:fermcut-off}
\end{equation}
We impose the boundary condition that the scattering amplitude in the physical
limit $k\rightarrow0$ reproduces the fermion-fermion scattering length,
\begin{equation}
\frac{1}{T(p)}=\frac{1}{g^{2}}\,\Pi(P_{0},P,0)=\frac{M}{4\pi a_{F}}\,.
\end{equation}
Here $P_{0}\ (P)$ denote the total energy (momentum) flowing through the system
and $p=\sqrt{2MP_{0}-P^{2}/2}$ is the relative momentum of the two fermions. 
Integrating the resulting ERG equation gives \cite{Bir}
\begin{eqnarray}
\Pi(P_{0},P,k) & =&\frac{g^{2}M}{4\pi^{2}}\biggl[-\frac{4}{3}\,k
+\frac{\pi}{a_{F}}+\frac{16}{3k}\left(MP_{0}-\frac{P^{2}}{2}\right)\cr
&& \qquad\qquad-\frac{P^{3}}{24k^{2}}+...\biggr].
\end{eqnarray}

Using a gradient expansion of the action, we can define boson wave-function
and mass renormalisation factors by
\begin{equation}
Z_{\phi}(k)=\frac{\partial}{\partial P_{0}}\,\Pi(P_{0},\vec{P},k)\Biggr|_{P_{0}
=\mathcal{E}_{D},\vec{P}=0},
\end{equation}
and
\begin{equation}
\frac{1}{4M}\,Z_{m}(k)=-\frac{\partial}{\partial P^{2}}\,\Pi(P_{0},\vec{P},k)
\Biggr|_{P_{0}=\mathcal{E}_{D},\vec{P}=0},
\end{equation}
where $\mathcal{E}_{D}=-1/(Ma_{F}^{2})$ denotes the bound-state energy
of a pair of fermions. Note that these renormalisation factors are
only identical in vacuum for a limited subset of cutoff functions
(which, like (\ref{eq:fermcut-off}) must preserve Galilean invariance
to lowest order), otherwise the identity $Z_{\phi}(k)=Z_{m}(k)$ holds
only in the physical limit $k\rightarrow0$ and the evolution of these
renormalisation factors should be calculated separately. 

The evolution of the boson-boson scattering amplitude follows from
\begin{equation}
-\frac{2}{(2\pi)^{4}}\partial_{k}u_{2}(\mathcal{E}_{D},k)
=\frac{\delta^{4}}{\delta\phi^{2}(\mathcal{E}_{D},0)
\delta\phi^{\dagger2}(\mathcal{E}_{D},0)}\,
\partial_{k}\Gamma|_{\phi=0}.
\end{equation}
This equation can be separated into fermionic and bosonic contributions
containing $\partial_{k}R_{F}$ and $\partial_{k}R_{B}$, respectively.
We first look at the mean-field result, where bosonic contributions
are neglected. The evolution of $u_{2}$ is then given by 
\begin{equation}
\partial_{k}u_{2}=-\frac{3g^{4}}{4}\int\frac{d^{3}{\vec{q}}}{(2\pi)^{3}}\,
\frac{\partial_{k}R_{F}}{\left[(E_{FR}(\vec{q},k)-\mathcal{E}_{D}/2\right)]^{4}},
\end{equation}
where $E_{FR}(\vec{q},k)=\frac{1}{2M}\, q^{2}+R_{F}(q,k)$. 
Integrating this gives
\begin{equation}
u_{2}(0)=\frac{1}{16\pi}\,M^{3}g^{4}a_{F}^{3},
\end{equation}
where we have again used the sharp cutoff function of Eq.~(\ref{eq:fermcut-off}).
The scattering amplitude at threshold is 
\begin{equation}
T_{BB}=\frac{8\pi}{2M}\,a_{B}=\frac{2u_{2}(0)}{Z_{\phi}^{2}}=\frac{8\pi a_{F}}{M},
\end{equation}
giving the well-known mean-field result $a_{B}=2a_{F}$ \cite{Ha}
which is far from the exact value $a_{B}=0.6a_{F}$ \cite{Shl}. This
implies that beyond-mean-field effects such as dimer-dimer rescattering
are important be considered. 

To include such effects we must take into account the boson loops. After some 
algebra, we find
\begin{equation}
\partial_{k}u_{2}|_{B}=\frac{u_{2}^{2}(k)}{2Z_{\phi}^{3}(k)}
\int\frac{d^{3}{\vec{q}}}{(2\pi)^{3}}\,
\frac{\partial_{k}R_{B}}{\left[E_{BR}(\vec{q},k)-\mathcal{E}_{D}\right]^{2}},
\end{equation}
where 
\begin{equation}
E_{BR}(\vec{q},k)=\frac{1}{4M}\, q^{2}+\frac{u_{1}(k)}{Z_{\phi}(k)}
+\frac{R_{B}(q,k)}{Z_{\phi}(k)}\end{equation}
 and \begin{equation}
u_{1}(k)=-\Pi(\mathcal{E}_{D},0,k).\end{equation}
We choose the bosonic cutoff function to be as close as possible to
the fermionic one, 
\begin{equation}
R_{B}(\vec{q},k)=Z_{\phi}\frac{(c_{B}k)^{2}-q^{2}}{4M}\,\theta(c_{B}k-q),
\end{equation}
apart from the addition of a parameter $c_{B}$, which sets the relative
scale of the fermionic and bosonic regulators, and a factor of $Z_{\phi}$.
The latter has the important advantage of leading to a consistent
scaling behaviour, so that all contributions to a single evolution 
equation decay with the same power of $k$ for large $k$. Moreover 
it also gives $a_{F}$-scaling, where all terms in a single equation have 
the same dependence on $a_{F}$.

The mean-field result is recovered for $c_{B}=0$ while the opposite limit  
$c_{B}\rightarrow\infty$ leads to $a_{B}\rightarrow0$. Using $c_{B}=1$ 
gives a ratio of $a_{B}/a_{F}=1.13$. Taking $c_{B}=\sqrt{2}$  as in 
Ref.~\cite{Diehl} results in $a_{B}/a_{F}=0.75$. In general, the results 
show a rather strong dependence on the relative scale parameter $c_{B}$, 
as can be seen below in Fig.~\ref{fig:Ratio}. Such dependence of a physical
result on the choice of cutoff must be an artifact of our truncation
of the action, since it should vanish for an exact solution of the 
functional RG equation. It indicates that the minimal ansatz used for 
the effective action needs to be extended. 

We now consider the effect of adding the local three-body force of
Eq.~(\ref{eq:ansatz3}). The evolution equations for $u_{1}$ and $Z_{\phi}$ 
remain unchanged but the one for $u_{2}$ now becomes
\begin{eqnarray}
\partial_{k}u_{2}&= & -\,\frac{3g^{4}}{4}\int\frac{d^{3}{\vec{q}}}{(2\pi)^{3}}\,
\frac{\partial_{k}R_{F}}{\left[E_{FR}(\vec{q},k)-\mathcal{E}_{D}/2\right]^{4}}
\nonumber \\
&& -2\lambda g^{2}\int\frac{d^{3}{\vec{q}}}{(2\pi)^{3}}\,
\frac{\partial_{k}R_{F}}{\left[E_{FR}(\vec{q},k)-\mathcal{E}_{D}/2\right]^{3}}
\nonumber \\
&& +\,\frac{u_{2}^{2}}{2Z_{\phi}}\int\frac{d^{3}{\vec{q}}}{(2\pi)^{3}}\,
\frac{\partial_{k}R_{B}}{\left[E_{BR}(\vec{q},k)-\mathcal{E}_{D}\right]^{2}}.
\label{eq:u2sym}
\end{eqnarray}
The evolution equation for $\lambda$ is defined by an expansion about
the energy of the bound state pole for bosons, and half that 
energy for fermions, 
\begin{equation}
\partial_{k}\lambda =-\frac{i}{2}\,\frac{\delta^{4}
\STr\left[\partial_{k}R(\Gamma^{(2)}-R)^{-1}\right]}
{\delta\phi^{\dagger}\!(\mathcal{E}_{D},0)
\delta\phi(\mathcal{E}_{D},0)\delta\psi^{\dagger}\!(\mathcal{E}_{D}/2,0)
\delta\psi(\mathcal{E}_{D}/2,0)}.
\end{equation}

\begin{figure}
\begin{centering}
\includegraphics[width=7cm]{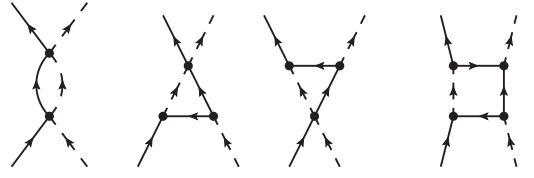}
\par\end{centering}
\caption{The skeletons of the diagrams that contribute to the evolution of
the three-body term. A dashed line denotes a boson, a solid line a
fermion. Each diagram can have one insertion of $\partial_{k}R_{F,B}$ as 
appropriate on a single internal line. We have the ladder diagram on 
the left, two triangle diagrams and finally the box diagram. 
With insertions there are 24 diagrams in total. \label{fig:skeleton}}
\end{figure}

There are three distinct contributions to the running of $\lambda$, coming 
from ladder, triangle and box diagrams, as shown in Fig.~\ref{fig:skeleton}.
We denote the corresponding driving terms as $D_{l}$, $D_{t}$ and
$D_{b}$, splitting the last two into their fermionic and bosonic 
contributions. After evaluation of traces and contour integrals, we get 
\begin{align}
D_{l} & =\lambda^{2}\int\frac{d^{3}{\vec{q}}}{(2\pi)^{3}}\,
\frac{\partial_{k}(R_{F}Z_{\phi})
+\partial_{k}R_{B}}{(E_{FR,D}Z_{\phi}+E_{BR,D})^{2}},\\
D_{t}^{F} & =g^{2}\lambda\int\frac{d^{3}{\vec{q}}}{(2\pi)^{3}}\,
\frac{\partial_{k}R_{F}(E_{BR,D}+2Z_{\phi}E_{FR,D})}{E_{FR,D}^{2}(E_{FR,D}Z_{\phi}
+E_{BR,D})^{2}},\\
D_{t}^{B} & =g^{2}\lambda\int\frac{d^{3}{\vec{q}}}{(2\pi)^{3}}\,
\frac{\partial_{k}R_{B}}{E_{FR,D}(E_{FR,D}Z_{\phi}+E_{BR,D})^{2}},\\
D_{b}^{F} & =\frac{g^{4}}{4}\int\frac{d^{3}{\vec{q}}}{(2\pi)^{3}}\,
\frac{\partial_{k}R_{F}(2E_{BR,D}+3Z_{\phi}E_{FR,D})}{E_{FR,D}^{2}(E_{FR,D}Z_{\phi}
+E_{BR,D})^{2}},\\
D_{b}^{B} & =\frac{g^{4}}{4}\int\frac{d^{3}{\vec{q}}}{(2\pi)^{3}}\,
\frac{\partial_{k}R_{B}}{E_{FR,D}^{2}(E_{FR,D}Z_{\phi}+E_{BR,D})^{2}},\end{align}
where $E_{FR,D}=E_{FR}-\mathcal{E}{}_{D}/2$ and 
$E_{BR,D}=E_{BR}-\mathcal{E}{}_{D}$.
Note that the evolution of the three-fermion interaction $\lambda$ does
not depend on the four-fermion interaction $u_{2}$, but $u_{2}$
does depend on $\lambda$. For the initial condition on $\lambda$, we demand
that there is no fundamental three-fermion interaction, and hence
$\lambda\rightarrow 0$ as $k\rightarrow\infty$. In practical calculations it
is simpler to take $\lambda$ to be zero at some large starting scale. Since 
$\lambda$ behaves like $1/k^{2}$ for large $k$, this can be used reliably.

Now we turn to the results. We note that these are numerically independent 
of the starting scale provided it is chosen to be at least $ka_{F}\simeq 100$. 
For $k\gg1/a_{F}$ the system is in the ``scaling regime'', and a fixed point
corresponding to the unitary limit governs the evolution until $k$ becomes 
comparable with $1/a_{F}$. (This can best be seen from the evolution of the
dimensionless coupling $k^{2}\lambda$.)

We find that the ratio $a_{B}/a_{F}$ decreases when the three-body term is 
included. For example, choosing $c_{B}=1$ leads to $a_{B}/a_{F}=0.74$, which 
should be compared to $a_{B}/a_{F}=1.13$ without the three-body term. 
Similarly the choice $c_{B}=\sqrt{2}$ \cite{Diehl} gives $a_{B}/a_{F}=0.69$. 

\begin{figure}
\begin{centering}
\includegraphics[clip,width=6.5cm]{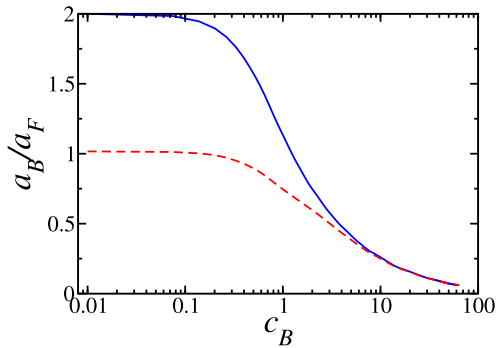} 
\par\end{centering}
\caption{Ratio of boson-boson to fermion-fermion scattering lengths as a 
function of the relative scale parameter $c_{B}$. The blue (solid) curve 
shows results for the minimal action $\Gamma_{\mathrm{min}}$; the red 
(dashed) curve shows the effect of adding the local three-body 
term. \label{fig:Ratio}}
\end{figure}

The full behaviour of $a_{B}/a_{F}$ as a function of $c_{B}$ is presented 
in Fig.~\ref{fig:Ratio}. This shows that, as well as reducing the overall
size of the ratio $a_{B}/a_{F}$, the inclusion of the three-body force 
significantly weakens its dependence on the relative scale $c_{B}$. 
We expect the qualitative
features of this picture to remain correct for any bosonic regulator 
although the quantitative details will depend on the particular functional 
form used.

Note that for large $c_{B}$, the dominant contributions to $a_{B}/a_{F}$ 
comes from the boson-loop terms in the equation for $u_{2}$. Since these 
do not depend on the three-body coupling $\lambda$, the two curves approach 
each other. Moreover, this limit corresponds to integrating out the 
fermions first, which generates a non-zero value for $u_2$ at the start of the 
bosonic integration. In the limit $c_B \rightarrow \infty$, this coupling is 
driven to the trivial fixed point, $u_2=0$, since we have no terms to cancel the 
linearly divergent boson-boson loop diagram and the diagrams with three-body 
couplings are too weak to alter this behaviour.

On the other hand, the main contributions for small $c_{B}$ come from the 
fermion and mixed fermion-boson loops, the latter arising in the three-body
coupling. In particular, the mixed boson-fermion loop diagrams containing 
the fermionic cut-off contribute to the evolution of the three-body coupling, 
even when the bosonic degrees of freedom have been integrated out. As a result, 
inclusion of $\lambda$ leads to a significant deviation from the mean-field 
result, $a_{B}/a_{F} = 2$, that survives in the limit $c_B \rightarrow 0$.

These results for very large or very small values of $c_B$ should not
be taken too seriously. Arguments based on ``optimisation'' of the
cut-off function, see Ref.~\cite{Pawlowski}, indicate that one should
choose the cut-off to try to maximise the rate of convergence for our
expansion of the action. Although a precise criterion has not yet been
defined for nonrelativistic theories, such arguments suggest a
choice where bosons and fermion cut-offs run at roughly the same rate,
i.e. $c_B$ is of the order of 1.

In spite of this clear improvement over calculations that include
two-body interactions only, adding the simplest possible three-body
term is not enough to ensure that the results are completely independent
of the parameter $c_{B}$ in the region $c_{B}\simeq 1$. It is worth 
emphasising again that, as long as any truncation of the effective action 
is made, the results will never be completely independent
of the choice of cutoff. It seems likely that further extensions of
the effective action will result in stability of the results with
respect to the variations of $c_{B}$ in wider region. Such extensions
could include both four-body interactions as well as energy and/or momentum
dependent three-body forces. Work along these lines is now in progress.

In summary, we have performed an ERG analysis of the boson-boson
scattering length in a system of nonrelativistic fermions. Our study 
indicates that, while a simple ansatz with only local two-body interactions
can yield results that are close to the exact value, these results
are sensitive to the value of a parameter controlling the relative 
scales of the fermionic and bosonic cutoffs. We show that the inclusion 
of a local three-body interaction brings the scattering length
closer to the exact value and significantly reduces its sensitivity to the 
relative scale parameter. 

One of the authors (BK) is supported by the EU FP7 programme (Grant
219533). BK and MCB thank the Institute for Nuclear Theory at the
University of Washington for its hospitality and the Department of
Energy for partial support during the completion of this work.

\end{document}